\documentclass[galaxies,article,accept,moreauthors,pdftex]{Definitions/mdpi} 

\firstpage{1} 
\pubvolume{1}
\issuenum{1}
\articlenumber{0}
\pubyear{2024}
\copyrightyear{2024}
\datereceived{} 
\dateaccepted{} 
\datepublished{} 
\hreflink{https://doi.org/} 

\Title{The seeding of cosmic ray electrons by cluster radio galaxies: a review}
\TitleCitation{Cosmic rays by radio galaxies}

\Author{Franco Vazza $^{1,2,3}$ and Andrea Botteon$^{2}$}
\AuthorNames{Franco Vazza et al.  }
\AuthorCitation{Vazza F., Botteon A.}
\address{%
$^{1}$ \quad Dipartimento di Fisica e Astronomia, Universit\'{a} di Bologna, Via Gobetti 93/2, 40129 Bologna, Italy\\
$^{2}$ \quad INAF-Istituto di Radio Astronomia, via Gobetti 101, 40129 Bologna, Italy\\
$^{3}$ \quad Hamburger Sternwarte, Universit\"{a}t Hamburg, Gojenbergsweg 112, 41029 Hamburg, Germany\\}
\corres{Correspondence: franco.vazza2@unibo.it}

\abstract{Radio galaxies in clusters of galaxies are a prominent reservoir of magnetic fields and of non-thermal particles, which get mixed with the intracluster medium. We review the observational and theoretical knowledge of the role of these crucial ingredients for the formation of diffuse radio emission in clusters (radio halos, relics, mini halos) and outline the open questions in this field.} 

\begin{document}

\section{Introduction}

Jets from radio galaxies can store up to the majority of their internal energy in the form of non-thermal components (relativistic particles
and magnetic fields), although the balance between thermal and non-thermal components may depend on the radio galaxy type, as well as on the interaction with the surrounding environment \cite[e.g.][]{2008MNRAS.386.1709C,2018MNRAS.476.1614C}. Given the very active dynamics of the intracluster medium (ICM), under the effect of the episodic accretion of gas and dark matter \cite[e.g.][]{bn99,do05,va11turbo,2017MNRAS.470..142S, 2023arXiv231106339A}, as well as of the powerful  feedback from radio galaxies themselves \cite[e.g.][]{chu04,2006MNRAS.373L..65H,gaspari12,2014ApJ...789...54L,2023Galax..11...73B}, it appears unavoidable that such interactions result into the fuelling of a prominent and potentially visible reservoir of magnetic fields and non-thermal particles in the ICM  \cite[e.g.][]{Volk&Atoyan..ApJ.2000,hardcastlecroston}. 

Recent deep radio observations are detecting complex morphologies of remnant plasma, injected by radio galaxies and in different stages of mixing with the surrounding environment \cite[e.g.][]{2017SciA....3E1634D,2018MNRAS.473.3536W,2020A&A...634A...4M,2022MNRAS.514.3466Q,brienza22}, and often in conjunction with diffuse $\sim \rm Mpc$ sized cluster radio sources.

In the case of "radio relics" (i.e. elongated and polarised radio sources that are typically co-located with merger shock waves \cite[e.g.][for a review]{2019SSRv..215...16V}) a pre-existing population of low energy relativistic electrons is  often required to explain the brightness of radio relics associated with weak $\mathcal{M} \leq 2.5$ shocks, in which case shock re-acceleration, rather than the injection of freshly accelerated electrons from the thermal pool, can be hypothesised  \cite[e.g.][]{ka12, 2013MNRAS.435.1061P,2020A&A...634A..64B,2021ApJ...914...73Z}. In  "radio halos" (i.e. centrally located, roughly spherically symmetric and unpolarised sources believed to trace turbulence in cluster with recent mergers  \cite[e.g.][for a review]{bj14}) fossil electrons are required to allow Fermi II turbulent re-acceleration to produce the observed level of radio emission, which also shows hints of curvature at high energies. Similar requirements have been proposed also to explain  "radio mini-halos" (smaller than radio halos and preferentially found in cool-core clusters, \cite[e.g.][]{2008A&A...486L..31C,2020MNRAS.499.2934R}) and also the pervasive emission recently detected at the extreme periphery of clusters and in-between pairs of interacting clusters of galaxies \cite[e.g.][]{2019Sci...364..981G,bv20,dejong22,shweta20, botteon22a2255,2022Natur.609..911C,bruno23a2142,beduzzi23,nishiwaki23}.

In this review, we wish to present the state of the art on the  existing observational evidences for the mixing of plasma ejected by radio galaxies with the ICM (Sec.\ref{observations}), and to give the status of our theoretical and numerical understanding of this process, which requires the understanding of the dynamics of the ICM across a wide range of spatial scales (Sec.\ref{theory}).

\section{Observations}
\label{observations}

\begin{figure*}
    \centering
    \includegraphics[width=0.95\textwidth]{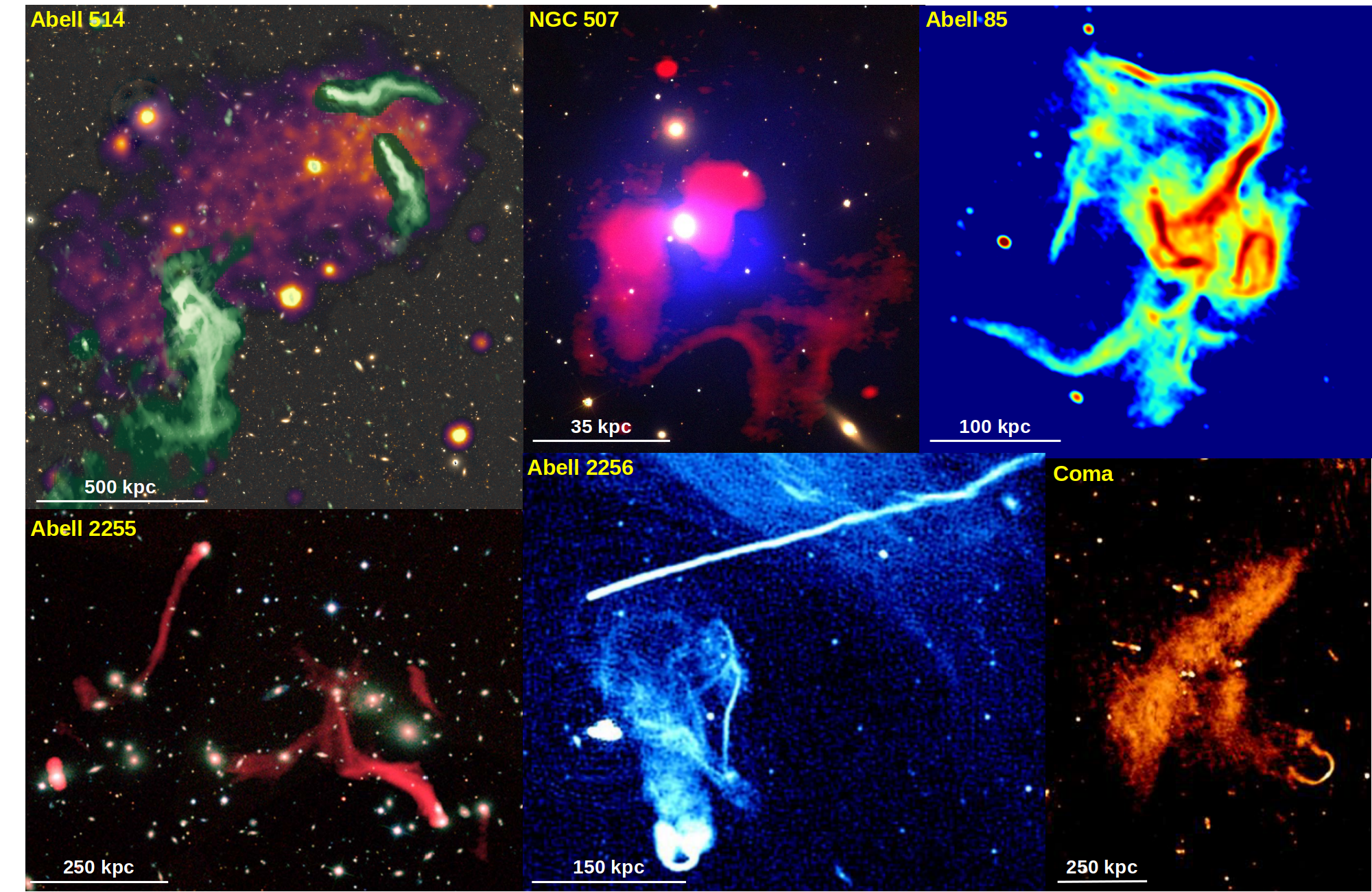}
   \caption{Examples of ongoing cosmic ray seeding and interaction between radio galaxies and thermal gas. From top left to bottom right: AGN with bended radio tails (green) due to the interaction with the X-ray gas (purple) in Abell 514 \citep{lee23a514}, radio plasma (red) transported by thermal gas (blue) sloshing motions in the NGC 507 group \citep{brienza22}, the filamentary radio phoenix in Abell 85 \citep{raja23arx}, the complex system of filamentary tailed radio AGN (red) at the center of Abell 2255 \citep{botteon20a2255}, the radio phoenix and the long head-tail radio galaxy in Abell 2256 \citep{rajpurohit23}, a radio galaxy feeding the radio relic in the Coma cluster \citep{bonafede22}.}
    \label{fig:gallery}
\end{figure*}

The electrons emitting in diffuse cluster radio sources  are, in most cases, expected not to be accelerated in-situ. 
Radio phoenixes and disturbed radio tails such as those shown in Fig.~\ref{fig:gallery} represent the clearest evidence of how the relativistic plasma ejected by cluster AGN can seed the ICM with non-thermal components, while being mixed and distributed in the surrounding environment by thermal gas motions. 

\subsection{Radio phoenixes}

Radio phoenixes are diffuse radio sources with sizes at most of a few hundred kpc, characterized by ultra-steep spectra ($\alpha > 1.8$, where $I(\nu) \propto \nu^{-\alpha}$ is the radio emision spectrum) showing high-frequency spectral steepening. They trace fossil radio plasma ejected from past episodes of AGN activity (also called "AGN remnant"), revived by the dynamical motions within the ICM. Consequently, such type of emission is also referred to as "revived fossil plasma". As a result of interactions with the thermal gas, phoenixes typically show amorphous morphologies, with  an often unclear connection with the host galaxy that originally ejected the radio plasma (cf. Fig.~\ref{fig:gallery}.). 

Despite radio phoenixes \footnote{Before the term "phoenix", introduced by Ref.~\cite{kempner04taxonomy}, these sources were called "relics". Nowadays, the term "radio relic" is used to classify large and arc-shaped diffuse emission associated with shock fronts in cluster outskirts} have been known for about three decades \citep[e.g.][]{slee84, slee98, slee01, subrahmanyan03, green04}, with the advent of high-quality observations, and particularly at  $\lesssim 1 \rm ~GHz$ frequencies (where the radio emission appears brighter due to its steep spectral index), in the latest years the number of these sources has significantly increased \citep[e.g.][]{mandal20, duchesne20a1127, botteon20a2255, botteon21a1775, hodgson21jellyfish, pandge21, pandge22, pasini22a1550, riseley22a3266, groenveld24sub}. Instances of phoenixes with extreme spectral indexes of $>$2.5 have also been found \citep[e.g.][]{slee01, subrahmanyan03, green04, mandal19, hodgson21jellyfish}. High-resolution images have started to reveal their complex morphologies, often characterized by filamentary structures \citep{juett13, owen14, werner16ophiuchus, mandal20, raja23arx}, see for example the remarkable cases of the phoenixes in Abell 85 and Abell 2256 reported in Fig.~\ref{fig:gallery}. The existence of these phoenixes indicates the presence of extended patches of old, relativistic electrons
in the sub-GeV energy range in the ICM. Although these electrons are fully relativistic (their a Lorentz factor is $\gamma \sim 100$) they are usually considered as low-energy, or event "mildly relativistic" electrons in this context, considering that the electrons emitting detectable radio emission have a $\geq \rm ~GeV$ energy
 (corresponding to $\gamma \gg 10^3$).
While statistical studies on the properties of radio phoenixes, such as their occurrence with cluster mass and dynamical state, are still lacking, the recent observations of large samples of clusters \citep[e.g.][]{vanweeren21, duchesne21eor, duchesne21palaeontology, botteon22dr2, hoang22, knowles22, duchesne24arx} suggest that sources of fossil plasma may be very common, if not ubiquitous, in all clusters of galaxies, especially at lower frequencies.

\subsection{Disturbed radio tails}

Jets originating from supermassive black holes residing in cluster AGN are deflected by the ram-pressure while their host galaxy moves through the ICM ($P_{\rm ram}\propto \rho v^2$, where $\rho$ is the ICM density and $v$ is the relative velocity between jets, or the galaxy, and the ICM), generating the so-called tailed radio sources \citep{miley80rev}. Depending on the degree of jet bending, these sources can be classified as wide-angle tails (WAT), narrow-angle tails (NAT), or head-tails (HT). The most spectacular tailed radio galaxies can extend beyond Mpc-scale, leaving a long trail of non-thermal plasma behind their path through the ICM. This material is eventually dispersed in the environment. 
Relativistic electrons in the magnetised ICM primarily loose their energy via synchortron and Inverse Compton losses, on the short timescale given by $t_{\rm cool} \sim \gamma / \dot \gamma \sim 10 \rm ~Myr$ (using $\gamma=10^4$, $B=1 ~\mu G$ and $z=0$ as reference values for the ICM) \citep[e.g.][]{sa99}. The cooling timescale approximately scales inversely with the energy of electrons ($t_{\rm cool} \propto 1/\gamma$), hence the least energetic electrons in tails are found at larger distances from the AGN, and steeper synchrotron spectra are progressively found along the tail. As a result, radio tails appear longer at lower frequencies.

However, new highly sensitive radio observations (again, mainly at low frequency), are revealing that tails (i) often exhibit longer extensions than what expected by the radiative lifetime of relativistic electrons \citep[e.g.][]{sebastian17b}, (ii) may feature regions of surface brightness and spectral index flattening (contrary to the gradual steepening expected by particle ageing) \citep{degasperin17gentle, cuciti18, edler22}, and (iii) frequently display disturbed morphology, especially at their terminal ends, where the tail structure "breaks" \citep[e.g.][]{vanweeren12a2256, wilber18a1132, srivastava20, vanweeren21, venturi22, botteon21a1775, ignesti22stormy, lusetti24tails}, see Fig.~\ref{fig:gallery}. These properties suggest a non-trivial ongoing interplay between the non-thermal components in tails and the surrounding thermal gas,  leading to processes that can sustain particles lifetime for periods of time and distances longer than usually expected. Radio tails thus naturally fuel the cluster environment with seed mildly relativistic electrons, and the latter can be re-energised already in this first interaction with the ICM, i.e. even before they get re-accelerated again by the processes leading to diffuse radio emissions (see next Section).

\begin{figure*}
    \centering
    \includegraphics[width=0.5\textwidth,height=0.4\textwidth]{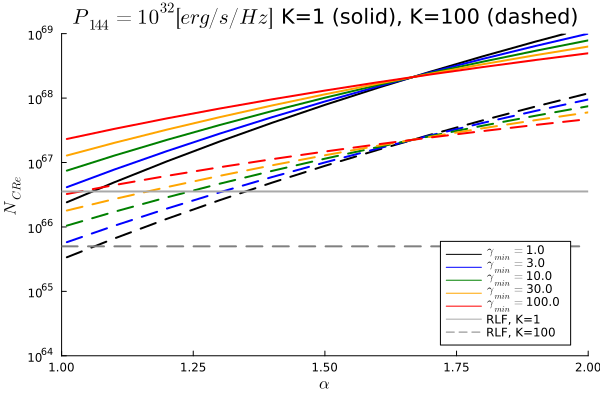}
       \includegraphics[width=0.55\textwidth,height=0.4\textwidth]{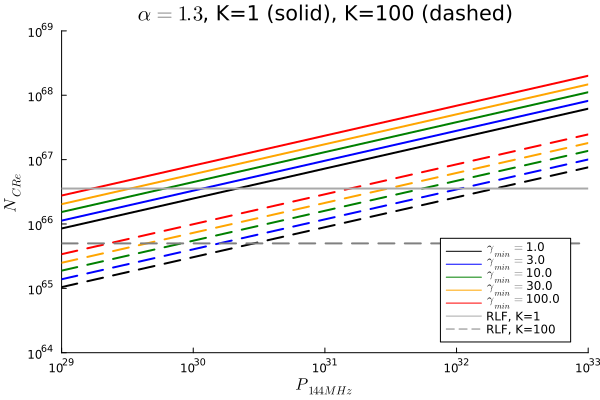}
   \caption{Top panel: number of cosmic ray electrons to explain a radio halo emitting $10^{32} \rm ~{erg/s/Hz}$ at 144 MHz, as a function of the observed radio spectrum ($\alpha$) and for different choices of the ratio between cosmic ray protons and electrons ($k$) and for different choices of the minimum energy of electrons.  Bottom panel: number of cosmic ray electrons required to explain radio halo emissions of different power at 144 MHz, assuming $\alpha=1.3$, $\gamma_{\rm min}=10$ and for $K=1$ or $K=100$. In both panels we assume that the radio halo has a volume of $1^3 \rm ~Mpc^3$ and that the host cluster has a $10^{15} ~M_{\odot}$ mass. The horizontal lines give the number of electrons injected by the entire population of radio galaxies in a cluster and only using a single activity burst,   assuming a $\alpha=0.8$ spectrum for all radio galaxies, and either $K=1$ or $K=100$. }
    \label{fig:NCRe}
\end{figure*}

\subsection{Evidence of interaction with the ICM and connection with large-scale diffuse radio emission}

Shock, cold fronts, turbulence and sloshing in the ICM can significantly influence the fate and morphology of non-thermal plasma ejected by cluster radio galaxies. Nowadays, radio and X-ray observations are providing clear evidence of this kind of interaction in many clusters (Fig.~\ref{fig:gallery}). Shocks or cold fronts co-located with sudden direction changes of radio galaxy tails have been reported in Abell 3411 \citep{vanweeren17a3411}, Coma \citep{lal20ngc4869}, Abell 3376 \citep{chibueze21}, Abell 1775 \citep{botteon21a1775}, Abell 3562 \citep{giacintucci22}, Abell 514 \citep{lee23a514}, and are invoked in other cases to explain the disturbed morphology of the radio emission \citep{pfrommer11, botteon19a781, wilber19, gendronmarsolais21}. 
Shocks can re-energize cosmic ray electrons both via re-acceleration (Fermi I processes) and adiabatic compression \citep[e.g.][]{ensslin01, ensslin02relics, markevitch05}. Shear flows inside cold fronts stretch and amplify magnetic fields \citep[e.g.][]{zuhone11, chibueze21}, possibly dragging the relativistic plasma with them in tangential directions \citep[e.g.][]{chibueze21, gendronmarsolais21, brienza22, giacintucci22, lee23a514, botteon24}. Turbulence, either induced by cluster-cluster mergers or by sloshing or by the wake of a fast-moving radio galaxy, can trigger Fermi II re-acceleration mechanisms \citep{pacholczyk76, wilson77, jones79, brunetti01coma, petrosian01}. In the extreme case of the so-called "Gently Re-Energized Tail" (GReET) in Abell 1033, a turbulent re-acceleration mechanism acting on a timescale comparable to that of relativistic electrons, barely balancing their cooling, has been proposed to explain the observed radio properties \citep{degasperin17gentle, edler22}. Adiabatic compression of fossil plasma due to a recent shock passage is also the favored formation scenario for radio phoenixes \citep{ensslin01}. However, as of now, the detection of a shock front at the location of a radio phoenix has been reported only in Abell 2443 \citep{clarke13}, leaving the formation scenario for these sources still uncertain (although the detection of shock-heated gas co-located with phoenixes has also been claimed in some cases \citep{ogrean11, schellenberger22, rahaman22a85, whyley24arx}).

Evidence of the seeding of radio halos and relics by radio galaxies has also been reported. Radio galaxies embedded into radio relics suggest the ongoing feeding of cosmic ray electrons in the environment \citep{bonafede14reacc, shimwell15, botteon16a115, vanweeren17a3411, digennaro18sausage, stuardi19, bonafede22}. In the case of Abell 3411, a clear connection between a radio galaxy compressed by a shock front and a radio relic has been used to support the scenario in which particles are re-accelerated by the weak shock \citep{vanweeren17a3411}, and test the adiabatic compression scenario \citep{button20}. Radio galaxies whose tails fade into radio halos or that demonstrate a complex mixing with central non-thermal diffuse emission/turbulent motions have been also associated with the seeding of cosmic rays on cluster scales \citep[e.g.][]{wilber18a1132, botteon20a2255, botteon22a2255, gendronmarsolais21, riseley22a3266, dejong22, velovic23}. Recently, a connection between a radio phoenix and central diffuse emission in Abell 85 has also been reported \citep{raja23}.

Although here we focus on systems with strong indications of interaction between AGN radio plasma and diffuse radio emission on larger scales, it is worth noting that several interesting observations detected different stages of the likely interplay between radio galaxies and the surrounding medium \cite[e.g.][to cite a few]{2020A&A...636L...1R, rudnick21a3266, condon21, 2021NatAs...5.1261B,2022ApJ...935..168R}. These interactions are often characterized by filamentary radio structures, which can in principle yield precious information about the life cycle of relativistic fossil electrons in the ICM.

Despite the observational evidence of the seeding of cosmic ray electrons by cluster radio galaxies, a number of questions still need to be addressed. What is the contribution of the single dynamical processes (shocks and turbulence) in the re-distribution and re-acceleration of cosmic ray electrons in clusters? How many radio galaxies are needed to fuel the ICM with non-thermal particles? Is a direct connection between radio galaxies and diffuse sources always needed to explain the emission from radio halos and relics? What about diffuse sources that do not show such a connection with cluster AGN?

\begin{figure*}
    \centering
    \includegraphics[width=0.40\textwidth]{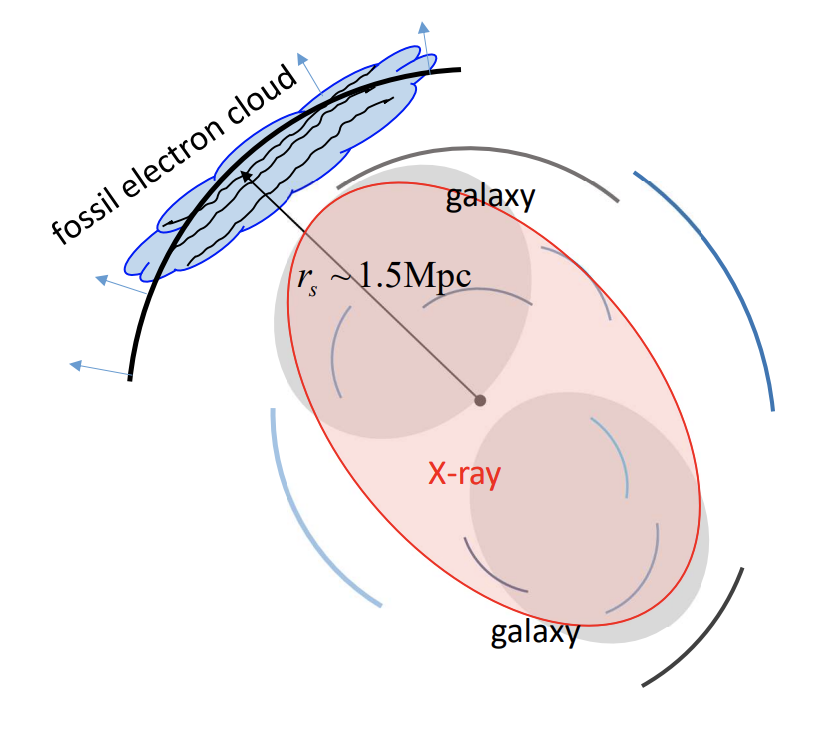}
     \includegraphics[width=0.52\textwidth]{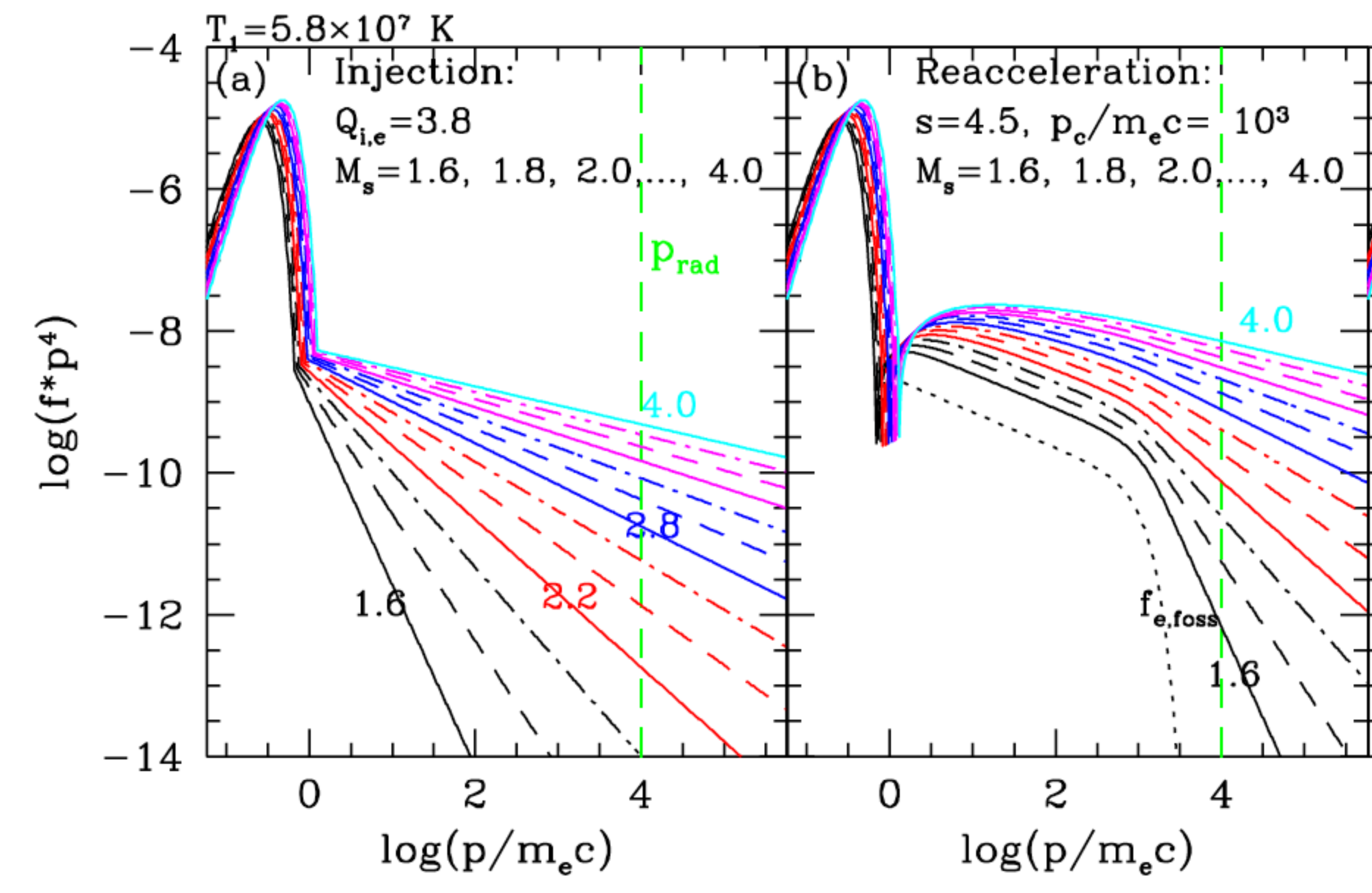}
   \caption{Left: cartoon sketch of the bubble of fossil relativistic electrons at the location of the "Sausage" radio relic in cluster CIZA J2242.8+5301. Right plots: electrons spectra resulting from the injection of fresh electrons via diffusive shock acceleration, as a function of Mach number (central panel), or from the re-acceleration of fossil relativistic electrons by the same shocks. Taken from \cite{2015ApJ...809..186K}.}
    \label{fig:kang}
\end{figure*}
\section{Theoretical and numerical models}
\label{theory}

Already one of the very first seminal works on this subject, Ref.~\cite{1977ApJ...212....1J}, noted that the contribution from a few radio galaxies hosted in the Coma cluster may explain the normalisation of the observed radio halo emission, provided that the magnetic field in the entire region was of order $\sim \mu G$. Similar estimates were later attempted for different systems, e.g. in A3562  \cite{2003A&A...402..913V}, or more recently for A2163 \cite{2020ApJ...897..115S}, just to cite a few.

A rough estimate can be obtained assuming energy equipartition between the relativistic electrons - also including those non emitting at the frequency of radio observations -  the ICM magnetic field and possibly the additional cosmic ray protons in the system (whose energy ratio with respect to electrons is $K$),  assuming a power-law distribution of cosmic ray electrons between $\gamma_{\rm min}$ and $\gamma_{\rm max}$~~($N(\gamma)=N_0 \gamma^{-\delta}$).
Here we follow the formalism by Ref.~\cite{1997A&A...325..898B}, who computed the equipartition magnetic field, $B_{\rm eq}$, also considering the contribution from cosmic ray electrons with energies below the observing frequencies, as:

\begin{equation}
B_{\rm eq}=[C(\alpha) \frac{P(\nu) \nu^{\alpha}}{\phi V}(1+K)]^{\frac{1}{\alpha+3}} {\gamma_{\rm min}}^{\frac{1-2\alpha}{\alpha+3}}
\end{equation}

\noindent where $P(\nu)$ is the radio emission (in units $\rm erg/s/Hz$), $\nu$ is the observing frequency (in $\rm Hz$),  $\alpha=(\delta-1)/2$ is the radio spectral index, $V$ is the volume of the emitting region (in $\rm cm^3$), and $\phi$ is the filling factor.  $C(\alpha)$ is a function of the spectral index, whose expression is given by Eq.~A3 in Ref.~\cite{1997A&A...325..898B}. 
The total energy of cosmic ray electrons (in units $\rm erg$), across their entire energy range, is thus given by: 

\begin{equation}
U_{\rm CRe} = \frac{2 \phi V}{\alpha+1}\frac{B_{\rm eq}^2}{(1+K)8 \pi }=m_e c^2 N_0 \int^{\gamma_{\rm max}}_{\gamma_{\rm min}} \gamma^{-\delta+1} d\gamma
\end{equation}

\noindent From the above relation, we derive the total number of electrons, which has the advantage of being a conserved quantity within the cluster volume (regardless of radiative losses or re-acceleration by turbulence). It should be noticed that the above estimate neglects the possibility of a continuous production of secondary electrons by the hadronic process \citep[][]{bl99}, which however cannot be the dominant process to explain radio halos, given the very tight constraints imposed by the non detection of hadronic $\gamma$-ray emission by high energy observations \citep[e.g.][]{alek10,fermi14,ackermann16}. 

The total number of cosmic ray electrons can be thus computed with: 

\begin{equation}
N_{\rm CRe} =  \frac{U_{\rm CRe}}{m_e c^2} \cdot \left(\frac{-\delta+2}{-\delta+1}\right) \cdot  \left(\frac{\gamma_{\rm min}^{-\delta+1}-\gamma_{\rm max}^{-\delta+1}}{\gamma_{\rm min}^{-\delta+2}-\gamma_{\rm max}^{-\delta+2}}\right).
\end{equation}

\noindent One can use the above relation, after specifying the minimum and maximum energies of the cosmic ray electron population, to estimate the number of electrons required by an observed radio structure for which energy equipartition can be reasonably hypothesised. 

The left panel of Figure ~\ref{fig:NCRe} gives the total number of cosmic ray electrons under different model variations (e.g. for different values of $K$, $\gamma_{\rm min}$) for a radio halo with volume $V=1^3 ~\rm Mpc^3$ at $z=0$ and with a typical radio power of $10^{32} ~\rm{ erg/s/Hz}$ at $144$ MHz, for different values of the radio spectral index. The right panel of the same Figure shows the number of electrons as a function of the radio power instead, by fixing the spectrum to a reference $\alpha=1.3$ value. 

The two additional horizontal lines in each panel mark the total number of cosmic ray electrons injected by all radio galaxies present in a $10^{15} M_{\odot}$ cluster (again for the $K=1$ or $K=100$ extreme scenarios, and always considering $\gamma_{\rm min}=10$). We computed this quantity based on the 
the same formalism outlined above, i.e. after assuming equiparition between magnetic fields and cosmic rays,  but this time using the distribution of radio luminosities of cluster radio galaxies derived at 843 MHz by Ref.~\cite{2017MNRAS.467.3737G}, and assuming a power-size relation from Ref.~\cite{2023A&A...677A.188B} to get the largest linear scale (LLS). From the latter, we estimate the volume occupied by radio lobes following the recent theoretical results by Ref.~\cite{2023MNRAS.526.3421S}, who studied the evolution of simulated FRI radio sources in idealised cluster atmospheres, reporting an approximate scaling $V_g \sim LLS^3 \cdot (\rm 10~kpc/LLS)$ between the size of expanding lobes and the volume they occupy. 

In essence, this simplistic model gives an idea of the number of cosmic ray electrons that can be available as a result of {\it a single generation} of radio lobes by all radio galaxies in a typical cluster. Notice that, since it is reasonable that the $K$ ratio in lobes and in radio halos is the same, to a first approximation, only the lines referred to an equal $K$ should be compared here.  Using as a reference the case of a $\alpha=1.3$ radio halo in a $10^{15} ~M_{\odot}$ cluster, the left panel shows that a single episode of activity by all radiogalaxies in a cluster can easily account for all electrons required by a $\sim 5 \cdot 10^{30}-10^{31} ~\rm erg/s/Hz$ halo at 144~MHz, while for more powerful halos multiple burst episodes (up to $\sim 10$) are required instead. The actual number of bursts from radio galaxies can be slightly lowered, if the additional injection by cosmic rays by galactic winds, or structure formation shocks is included \cite[e.g.][]{volk99,2018ApJ...856..112F}. This simple formalism can also be extended to other kinds of diffuse sources like radio relics, yet the assumption of equipartition between cosmic rays and magnetic fields is much more questionable, owing to the much shorter acceleration timescale that is supposed to be at work in shock-accelerated cosmic rays. 

Although this analysis suggests that the present and past activity by a typical population of radiogalaxies in the ICM  can explain the total number of electrons required to power diffuse radio sources, the actual ICM dynamics leading to the transport of cosmic rays has to be properly modelled, in order to test whether the injection by radio galaxies is a viable seeding scenario to quantitatively explain observations.

\begin{figure*}
    \centering
    \includegraphics[width=0.95\textwidth]{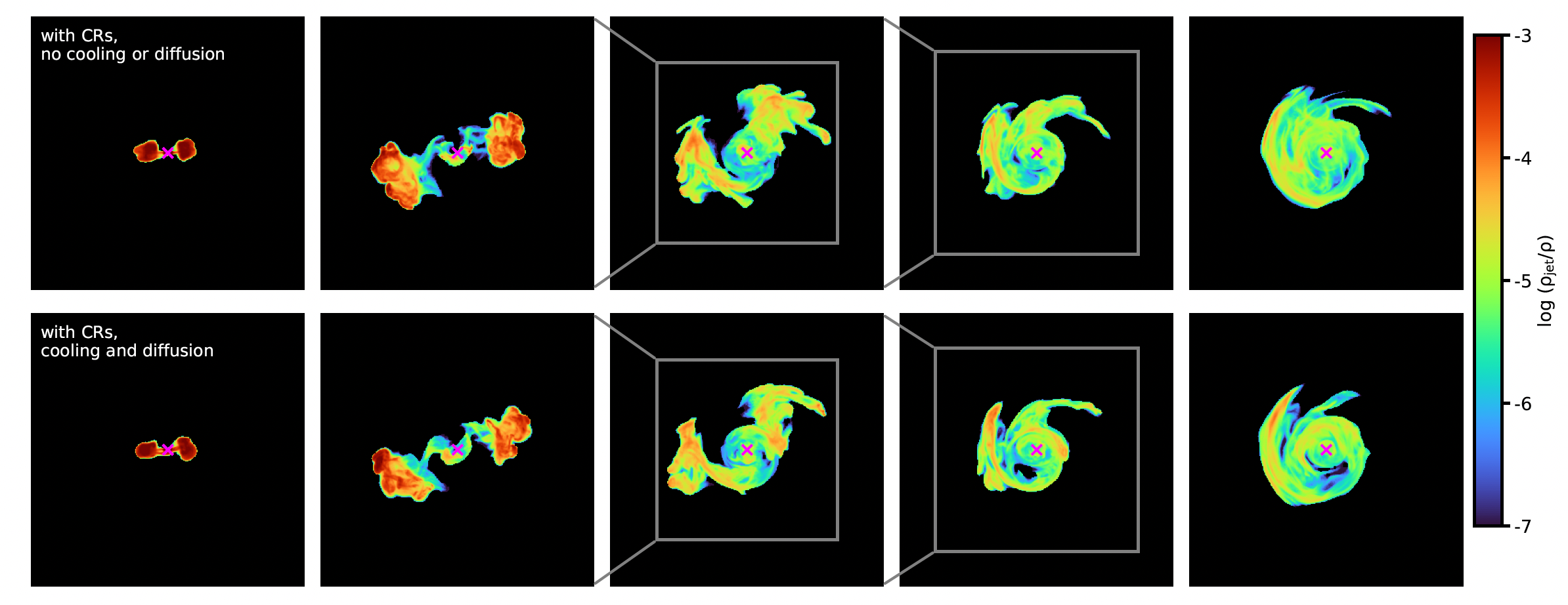}
    \includegraphics[width=0.98\textwidth]{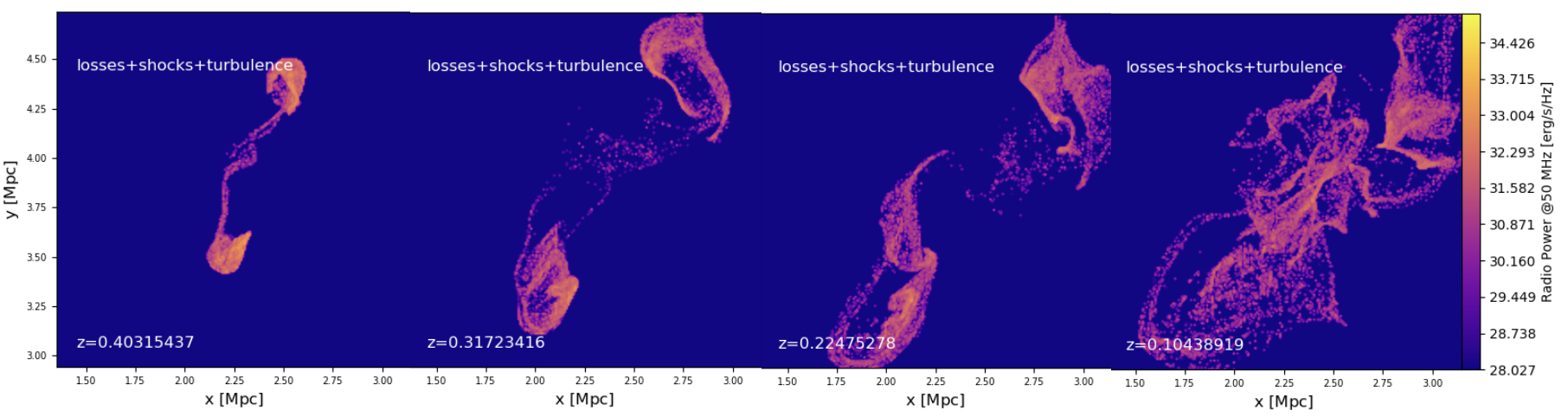}
   \caption{Top panels: Evolution of the projected density of fluid tracers ejected by AGN inflated bubbles, in simulations by Ref.~\cite{2021Galax...9...91Z}. The top row shows the evolution in the case in which cosmic rays are passively advected with the fluid, while in the lower row the effect of  cosmic rays diffusion and Alfv\'{e}n losses are included. Last row: Projected synchrotron radio emission at $150~\rm MHz$ and for four different evolutionary steps, for a simulated pair of AGN inflated bubbles in the cosmological simulation by Ref.~\cite{va21b}, in which electrons could age by radiative processes, as well as be re-energised by shocks and turbulent re-acceleration.}
    \label{fig:zuhone_vazza}
\end{figure*}

\subsection{Numerical simulations}

The actual transport of electrons injected by radio jets mixing with the (often turbulent) ICM can only be followed with fluid simulations, where also larger scale mixing motions induced by accretions, or by the repeated activity of AGN feedback, are included.  While accounting for the full multi-scale complexity of this process is a challenge even for modern numerical simulations, several works could at least capture the most important pattern associated with large-scale circulation of the material ejected by jets.

Starting from the pioneering work by Ref.~\cite{2002MNRAS.331..545B}, who simulated the mixing of buoyant bubbles in a typical cluster environment and studied their radio detectability as a function of time, numerical simulations have kept refining the state of the art of our understanding of how AGN feedback can affect the ICM \cite[e.g.][for a recent review]{2023Galax..11...73B}. Relatively fewer works focused on the interplay between the remnant plasma ejected by AGN and the cluster environment, and often resorting to a combination of different numerical methods in order to best capture the multi-scale nature of this process. 
For example, Ref.~\cite{2012ApJ...750..166M} used an Eulerian Magneto Hydrodynamical approach to account for "cluster weather" on radio lobes expanding into the perturbed environment of a cluster of galaxies produced by a  Smoothed Particle Hydrodynamics simulation. Likewise, the semi-analytic modelling of radio jets by Ref.~\cite{turner23} has been recently coupled to the Eulerian hydro dynamical resimulation of jets crossing the ICM extracted from a large suite of SPH cosmological simulations by Ref.~\cite{2023PASA...40...14Y}, which allowed them to probe the impressive resolution of $0.05 ~\rm kpc/cell$, and even to explore the potential connection with the production of  "odd radio circles" in follow-up work \cite{2024arXiv240209708S}.

A few works explored the possibility of a more direct connection between observed radio relics and the fuelling by single, nearby radio galaxies, by simulating the evolution of re-accelerated fossil electrons. 
In a series of works,Ref.~\cite{ka12}, \cite{2015ApJ...809..186K} and \cite{2018JKAS...51..185K}  modelled
the formation of radio relics using 1D diffusive-convection simulations with the goal of reproducing, for first, the integrated radio emission spectra and the profile of radio emission across the width of real relics (e.g.  "Toothbrush" in the merging cluster 1RXS J060303.3, and the "Sausage" in cluster CIZA J2242.8+5301). In particular, they tested direct shock acceleration models, based on diffusive shock acceleration, versus models in which the radio emitting particles were re-accelerated by weak shocks  ($\mathcal{M} \leq 2$) running over a pre-existing population of fossil electrons with a low-energy cutoff ($\gamma \sim 8 \cdot 10^4$). 
The latter scenario was shown to reproduce reasonably well the observed integrated spectrum, the profile of radio surface brightness for the Toothbrush as well as the downstream curvature of the spectrum.  A necessary ingredient for the model was therefore the presence of a wide and rather uniform distribution of "fossil" relativistic electrons, which they suggest could have been seeded by the previous activity of AGN in this sector of the cluster (e.g. Fig.~\ref{fig:kang}).



The formation and long-term evolution of disturbed radio tails in clusters  have been extensively investigated in a series of works by Ref.~\cite{2017PhPl...24d1402J,nolting19a,nolting19b,nolting23}, in which non-cosmological high resolution MHD simulations were coupled to the simulation of the cosmic ray fluid associated with remnant plasma. 

Ref.~\cite{2021ApJ...914...73Z, 2021Galax...9...91Z} investigated whether the morphology of some radio relics can reflect the distribution of fossil relativistic electrons previously injected by central AGN in clusters, and later dispersed into the ICM by sloshing gas motions generated by a previous merger event. They found that the transport of electrons along a spiral path, and its stretching tangentially to the spiral pattern, could produce filamentary emission pattern compatible with real radio relics (see e.g. Fig.\ref{fig:zuhone_vazza}). Following this approach, Ref.~\cite{botteon24} proposed a similar scenario to model the filamentary diffuse radio emission in Abel 2657, out of a shredded AGN bubble. 

Recently, Ref.~\cite{chibueze21} produced 
high-resolution MHD simulations to model the complex radio structure in  A3376, in which they accounted for the interaction between an AGN jet and a curved ICM magnetic field. They showed that the simulated jet propagation is quenched by the magnetic tension, which induces a later escape of the flow and of the relativistic particles carried with it, which can reasonably mimic the sharp  “double-scythe" shape of the real observation. The magnetic field which best reproduces the synchrotron morphology is of order $\sim 10 ~\mu G$, and the modelling suggests that magnetic reconnection might additionally be at work to produce an efficient injection of cosmic rays in the region.

However, whether a single burst from AGN, combined with turbulence and mixing motions in clusters can provide a sufficient seeding of relativistic electrons to cover the entire extent of observed radio relics, is not a fully settled issue.

To explore this question,  in a series of works, Ref.~\cite[][]{va21b,va23a} modelled the dynamics of radio jets mixing with the turbulent ICM in cosmological simulations, while also following the  spatial and spectral energy evolution of fossil electrons injected by AGN sources (see bottom panels of Fig.\ref{fig:zuhone_vazza}). Their simulation showed that 
right after the active stage of radio jets, the buoyant stage of mixing of relativistic electrons is soon followed by a longer stage of large scale mixing, driven by turbulent motions in the ICM. 
During that phase, the remnant radio plasma can remain detectable at low frequencies out to  $\leq 0.5$ Mpc distance from sources,  and even up to $\sim 0.5-1$ Gyr since its first injection, provided that re-acceleration from weak shocks or turbulence is triggered by the cluster accretion dynamics.  The emission from re-accelerated fossil electrons can produce detectable emission at any time, but it can typically produce only small  ($\leq 100$ kpc) and often filamentary radio features with steep radio spectra ($\alpha \geq 1.5-2$). 
 With the same approach, \cite{va23b} studied the mixing of electrons injected by eight realistic radio sources all activated at once in the same cluster, and compared their efficiency in seeding the ICM with cosmic rays to the one of shocks driven by mergers in the same cluster, finding that the latter are more naturally generating a volume filling distribution of fossil electrons on the scales typically covered by radio relics, as also reported by other complementary analysis \cite{inchingolo22,2023MNRAS.526.4234S}.
 

\begin{figure*}
    \centering
    \includegraphics[width=0.45\textwidth]{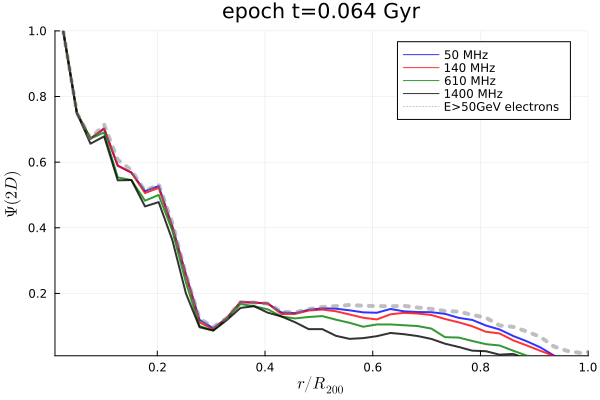}
    \includegraphics[width=0.45\textwidth]{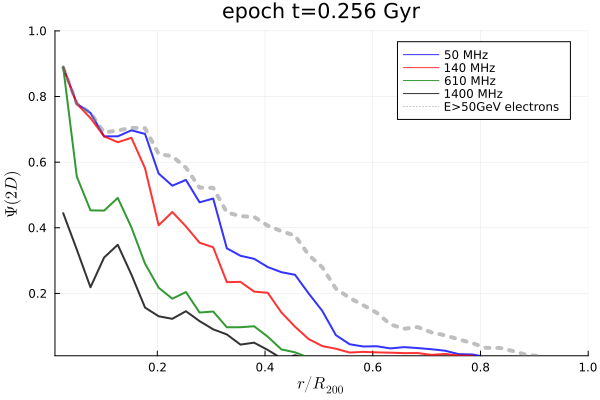}
    \includegraphics[width=0.45\textwidth]{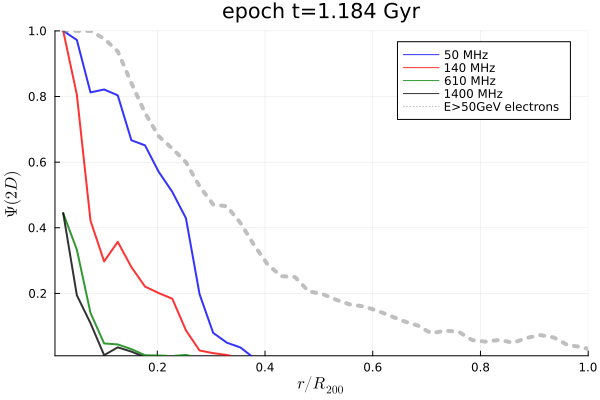}
   \caption{Area filling factors of detectable radio emission as a function of frequency (different colors) and of relativistic electrons (dotted gray lines) for three different epochs since the injection of radio jets in the cosmological simulation by \cite{va23b}.}
    \label{fig:filling}
\end{figure*}

A common prediction from all aforementioned numerical approaches is that the detectable structures associated with plasma tails or remnant plasma from radio galaxies can only probe "the tip of the iceberg" of a broader and more volume filling distribution of fossil electrons.
To visualise this, we show in Fig.~\ref{fig:filling} the measured evolution of the area filling factor, $\Psi(2D)$, of the radio emission from the same  simulation by \cite{va23b}, where eight radio galaxies release magnetic jets in a $\sim 10^{14}~ M_{\odot}$ galaxy cluster, all at $z=0.5$ and with a range of realistic jet powers. 
In the Figure, we show the fraction of area of radial shells from the cluster centre which is visible at different radio frequencies (50, 140, 610 and 1400 MHz), assuming for simplicity a detection threshold which scales with frequency as $\sigma(\nu)=\sigma_{LBA} [\nu/(50 ~\rm MHz)]^{-1}$, where $\sigma_{LBA}=\sigma = 5.7 \cdot 10^{-4} \rm ~Jy/beam$, as in \cite[e.g.][]{botteon22a2255}, assuming a beam of $\theta=12.5"$ and a (non evolving)  luminosity distance of the cluster of  $d_L=132 \rm ~Mpc$. The additional grey dotted lines in each panel show instead the area covering factor of the entire distribution of cosmic ray electrons injected by simulated radio galaxies, projected along the same line of sight. 
Of course, while the area filling factor of electrons mixing in the ICM can be large, the filling factor of the detectable radio emission can be much smaller, because it depends on the fraction of relativistic electrons in the high energy tail of the distribution, as well as on the observing frequency, and it varies with the available sensitivity of each radio observation. In particular, given the same volume filling distribution of relativistic electrons released by radio galaxies (or other mechanisms), different scenarios for the energy evolution of such electrons can vary their observable area filling factor at different radio wavelengths.
As it can be seen, the area filling factor of the observable emission rapidly drops with time, at an increased rate for higher frequencies. After  $\sim 1.2 \rm ~ Gyr$ since their injection in the cluster, the electrons from all galaxies have spread to large radii, 
due to the combined effect of their initial jet velocity of the large scale turbulent motions in the cluster, which can reach $\sigma_v \sim 300 ~\rm km/s$ in this simulated cluster \cite{va23a}. However,  cosmic rays dilute as they expand into the cluster volume and after $\sim 0.25 \rm ~Gyr$ since the injection, only less than $\sim 30\%$ of the projected area at $\sim 0.5 R_{200}$ is covered by cosmic rays, which further declines to $\sim 20\%$ one $\rm Gyr$ later. At this epoch, less than the $\sim 5\%$ of the projected area at $\sim R_{\rm 200}$ is covered by cosmic rays along the line of sight. 
The area covering factor of the observable emission is sharply dropping with frequency and time:  already $\sim 0.25 \rm ~Gyr$ since the jets injection, 
$\sim 30\%$ of the surface at $\sim 0.5 ~R_{200}$ is detectable at $50 \rm ~MHz$, but less than $\sim 5\%$ at $1400 \rm ~MHz$. 
One $\rm Gyr$ later, there is virtually no detectable emission from the remnant plasma from jets beyond $\sim 0.4 ~R_{200}$, and the very steep spectral energy distribution makes a detection only possible for low frequency (  $\leq 140 \rm ~MHz$) radio observations within that radius. 
It should be noted that this modelling cannot capture the additional effect of cosmic ray diffusion in the tangled ICM magnetic field, which can surely increase the volume filling factor of cosmic rays.  However, the timescales of the spatial diffusion of cosmic ray electrons are much longer than the evolutionary range considered here. In the range of energy of interest here, the diffusion coefficient should be in the  $D \leq 10^{30} \sim 10^{31} \rm ~cm^2/s$ ballpark, \cite[e.g.][and discussion therein]{bj14},  which means that the diffusion timescale over a $L \approx 1 \rm ~Mpc$ scale is $\tau_{\rm diff} \sim L^2/(4D) \geq 7.5-75 \rm ~Gyr$, and hence is a a very slow and sub-dominant mechanism. 

In summary, an in line with the simplistic theoretical derivation of the previous Section, the repeated activity of a few radio galaxies in clusters ($\leq 10$), joined with the typical turbulent gas motions expected from numerical simulations,  appears sufficient to  fill most of the area covered by observed radio halos ($\leq 0.1~R_{200})$ with relativistic electrons.

On the other hand, the external regions where radio relics typically form ($\geq 0.5 ~R_{\rm 200}$, e.g. \cite{2012MNRAS.421.1868V,2017MNRAS.470..240N}) are hardly replenished by significant amount of fossil cosmic ray electrons even by multiple radio galaxies, and probably 
additional cluster wide motions (e.g. sloshing) are required to conveniently distribute the  remnant plasma in a more uniform way, before shocks revive it, as already suggested \cite[e.g.][]{2021ApJ...914...73Z}.  How common such configurations may appear in the lifetime of cluster of galaxies remains to be assessed. Complementary to this,  the injection of cosmic rays by merger shocks can naturally lead to large and correlated populations of fossil electrons that subsequent re-acceleration events can illuminate on $\sim \rm ~Mpc$ scales \cite[][]{2013MNRAS.435.1061P,va23b,inchingolo22,2023MNRAS.526.4234S}, and moreover the ubiquitous presence of turbulent motions even in the outskirts of cluster can maintain electrons at a higher energy than expected only considering the effect of radiative losses, i.e. up to $\gamma \sim 10^3$ \cite{beduzzi23}.  This would imply the existence of a second and rather uniform population of fossil electrons on larger cluster scales, in addition to the outcome of evolving radio cluster galaxies.

\section{Conclusions and future perspectives}


In summary, both the direct observations of remnant radio plasma and of disturbed radio tails, and the more indirect indications of a connection between the activity of radio galaxies and large diffuse radio emissions in the clusters of galaxies (i.e. radio halos and radio relics) support the idea that the fuelling of relativistic electrons from cluster radio galaxies is a key  process to regulate the non-thermal energy budget of the ICM. 

From the theoretical viewpoint, the time-integrated activity of the population of radio galaxies in clusters appear sufficient to fuel the entire extent of diffuse radio emissions with the necessary amount of fossil electrons, while the exact timescales and efficiency with which relativistic electrons spread across the ICM depend on the multi-scale gas dynamics, and its investigation with modern numerical methods has just begun. Observations at very low-frequencies (< 100 MHz) performed with LOFAR2.0 and SKA-LOW will soon have the possibility to recover the emission from older populations of cosmic ray electrons in the ICM. In parallel, with the excellent polarisation capabilities of the new generation of radio telescopes (e.g. MeerKAT+, SKA-MID), these instruments will allow to investigate highly inefficient re-energization processes sustaining particles lifetime in the ICM.

Among the main processes to re-accelerate fossil electrons in the ICM, shocks and turbulence are the dominant mechanisms. While the detection of shock fronts has been already enabled by instruments such as \textit{Chandra}, \textit{XMM-Newton}, and \textit{Suzaku}, the direct observation of turbulent motions is limited to the case of Perseus by Hitomi \cite{hitomi16}. The deployment of new X-ray facilities capable of measuring the gas turbulent motions, either with large integration area (e.g. \textit{XRISM}) or in a more spatially resolved manner (e.g. \textit{Athena}), promises to enable a better constraining on the multi-scale dynamical interactions between the relativistic plasma dispersed by radio galaxies, and the evolving ICM.

\vspace{6pt}

\authorcontributions{F.V. \& A.B. curated the manuscript, the figures and the interpretation of results in equal parts. }


\conflictsofinterest{The authors declare no conflict of interest.}

\funding{F.V. acknowledges partial financial support from  the Cariplo ``BREAKTHRU'' funds Rif: 2022-2088 CUP J33C22004310003. A.B. acknowledges financial support from the European Union - Next Generation EU.}

\institutionalreview{Not applicable.}

\informedconsent{Not applicable.}

\dataavailability{Not applicable.}

\acknowledgments{The authors thank Annalisa Bonafede, Marisa Brienza, Hyesung Kang, Wonki Lee, Ramij Raja, Kamlesh Rajpurohit and John ZuHone for the kind permission to reuse the figures of their papers for this work. We thank Lawrence Rudnick for his constructive feedback on the manuscript, as well as our anonymus reviewers.}

\newcommand{\apss}{APSS}
\newcommand{\apjs}{ApJS}
\newcommand{\apj}{ApJ}
\newcommand{\aj}{ApJ}
\newcommand{\apjl}{ApJ Letters}
\newcommand{\mnras}{MNRAS}
\newcommand{\aap}{A\&A}
\newcommand{\ETC}{et al.}
\newcommand{\ssr}{Science \& Space Review}
\newcommand{\na}{New Astronomy}
\newcommand{\nar}{New Astronomy Review}
\newcommand{\nat}{Nature}
\newcommand{\physrep}{Physics Reports}
\newcommand{\aapr}{The Astronomy and Astrophysics Review}
\newcommand{\jcap}{Journal of Cosmology and Astroparticle Physics}
\newcommand{\pasp}{Publication of the Astronomical Society of the Pacific}
\newcommand{\sovast}{Soviet Astronomy}
\newcommand{\prd}{Physical Review Letters}
\newcommand{\jgr}{Journal of Geophysical Research}
\newcommand{\pasj}{Publications of the Astronomical Society of Japan}
\newcommand{\araa}{Annual Review of Astronomy and Astrophysics}
\newcommand{\prl}{Physical Review Letters}
\newcommand{\pre}{Physical Review Letters E}
\newcommand{\pasa}{Publications of the Astronomical Society of Australia}
 
\abbreviations{The following abbreviations are used in this manuscript:\\
\noindent
\begin{tabular}{@{}ll}
AGN & Active Galactic Nucleus\\
AMR& Adaptive Mesh Refinement\\
ASKAP& Australian Square Kilometre Array Pathfinder\\
BCG& Brightest Central Galaxy\\
DSA&Diffusive Shock Acceleration\\
GREET&Gently Re-Energized Tail\\
HBA& High Band Antenna\\
HT& Head Tail\\
ICM& Intra Cluster Medium\\
LBA& Low Band Antenna\\
LLS& Largest Linear Scale\\
LOFAR& Low Frequency Array\\
MHD& Magneto Hydro Dynamics\\
MWA& Murchison Widefield Array\\
NAT& Narrow Angle Tail\\
SMBH& Super Massive Black Hole\\
SPH& Smoothed Particle Hydrodynamics\\
WAT& Wide Angle Tail\\
$\Lambda$CDM& Lambda Cold Dark Matter
\end{tabular}}


\reftitle{References}

\externalbibliography{yes}
\bibliography{franco3,dan,library,arxiv}

\end{document}